\documentclass[referee]{raa}           
\usepackage{graphicx,times}
\usepackage{natbib}
\usepackage{amssymb,amsmath}
\bibpunct{(}{)}{;}{a}{}{,}

\usepackage[pagebackref=true]{hyperref}

\usepackage[bottom = 0.5in]{geometry}

\newcommand{\fermi}{{\it Fermi}}
\newcommand{\gr}{$\gamma$-ray}

\begin{document}

   \title{Identifying Three New AGNs Among {\it Fermi} Unidentified Gigaelectronvolt Sources}

 \volnopage{ {\bf 20XX} Vol.\ {\bf X} No. {\bf XX}, 000--000}
   \setcounter{page}{1}

   \author{Shun-Hao Ji\inst{1}, Zhong-Xiang Wang\inst{1,2}, Qiang-Meng Huang\inst{1}, Ruo-Heng Yang\inst{1}
   }

   \institute{ Department of Astronomy, School of Physics and Astronomy, Yunnan University, Kunming 650091, China; {\it jishunhao@mail.ynu.edu.cn, wangzx20@ynu.edu.cn}\\
        \and
             Shanghai Astronomical Observatory, Chinese Academy of Sciences, 80 Nandan Road, Shanghai 200030, China\\
\vs \no
   {\small Received 20XX Month Day; accepted 20XX Month Day}
}

\abstract{We report our identification of three gigaelectronvolt $\gamma$-ray
sources, 4FGL J0502.6+0036, 4FGL J1055.9+6507, and 4FGL J1708.2+5519, as
Active Galactic Nuclei (AGNs). They are listed in the latest 
\fermi-LAT source catalog as unidentified ones. We find that the sources all
showed $\gamma$-ray flux variations in recent years. Using
different survey catalogs, we are able to find a radio source
within the error circle of each source's position. Further analysis of optical 
sources in the fields allows us to
determine the optical counterparts, which showed similar variation patterns to
those seen in $\gamma$-rays. The optical counterparts have reported
redshifts of 0.6, 1.5, and 2.3, respectively, estimated from photometric 
measurements. In addition, we also obtain an X-ray spectrum of 4FGL J0502.6+0036 and a flux upper limit on the X-ray emission of 4FGL J1055.9+6507 by analyzing the archival data. The broadband spectral energy
distributions of the three sources from radio to $\gamma$-rays are 
constructed. Comparing mainly
the $\gamma$-ray properties
of the three sources with those of different sub-classes of AGNs, we tentatively
identify
them as blazars. Followup optical spectroscopy is highly warranted for
obtaining their spectral features and thus verifying the identification.
}

   \authorrunning{Ji et al. }            
   \titlerunning{Identifying three new AGNs among unidentified GeV sources}  
   \maketitle

%
\section{Introduction}           
\label{sect:intro}

Active Galactic Nuclei (AGNs) are powered by accretion onto super-massive black 
holes (SMBHs) in the centers of galaxies. Approximately 10\% of AGNs have radio 
jets, and the jet formation is widely considered due to two mechanisms: 
the Blandford-Znajek mechanism (\citealt{Blandford-z+1977}), in which the jet 
extracts the rotational energy of an SMBH, and the Blandford-Payne mechanism 
(\citealt{Blandford-p+1982}), in which the jet extracts the rotational energy 
of an accretion disk surrounding the SMBH. Depending on our viewing angle of
AGNs with jets, or so-called radio-loud AGNs, they can be divided into two 
classes (e.g., \citealt{Urry+95}). The first are blazars when our viewing
angles of their jets are small (generally $\leq 10^{\circ}$). Blazars are 
further divided into
two sub-types, BL Lac objects (BL Lacs) and Flat Spectrum Radio Quasars 
(FSRQs). BL Lacs have very weak or no emission lines in their optical 
emission, 
while FSRQs show emission lines with equivalent widths greater 
than 5\,\AA\ (\citealt{Urry+95}; \citealt{Scarpa+1997}). 
Due to the small viewing angles and thus strong Doppler boosting effect,
blazars are bright and highly variable sources in the sky.
The second class of radio-loud AGNs are misaligned ones (MAGNs), of which
we have large viewing angles. They tend to be fainter and usually closer to us
than blazars. MAGNs are further classified into radio galaxies (RGs) and Steep 
Spectrum Radio Quasars (SSRQs). 

The Large Area Telescope (LAT) onboard {\it the Fermi Gamma-ray Space 
Telescope (Fermi)} has been monitoring the high-energy $\gamma$-ray sky since 
2008 (\citealt{Atwood+09}). There are 7194 $\gamma$-ray sources in energy range
of from 50\,MeV to 1000\,GeV listed in the updated fourth source catalog 
(4FGL-DR4; \citealt{Ballet+23}). Nearly 66\% of them can be associated with 
a known astrophysical object, and among the associated ones, more than 80\%
are either blazars or MAGNs, while with 55 being the latter.
These AGN sources offer a good sample for various studies, for example, 
the contributions of blazars and MAGNs to the 
extragalactic $\gamma$-ray background (e.g., \citealt{Ajello+15}; \citealt{Ackermann+16}; \citealt{DiMauro+18}; \citealt{DiMauro+14}; \citealt{Fukazawa+22}),
the blazar sequence (e.g., \citealt{Ghisellini+08}; 
\citealt{Ghisellini+17}; \citealt{Fan+17}), strong variabilities 
(e.g., \citealt{Ghisellini+13}; \citealt{Hayashida+15}; \citealt{Shukla+18}; 
\citealt{Patel+18}).
For the other 34\% unidentified sources, blazars should dominate among them
and there may be a small number of MAGNs, according to the proportions of 
the associated sources.
Identification of these sources can help build the \gr\ sample as complete as
possible, thus helping probe the full physical properties of AGNs.
A popular way for source identification is through statistical analysis
or machine-learning methods 
(e.g., \citealt{Yi+17}; \citealt{Chiaro+21}; \citealt{Kaur+23}; \citealt{Zhu+24}), and by 
focusing on variations of optical light curves (R. Yang et al., in preparation),
we have tested to build our machine-learning method. In addition,
we are also carrying out identification of the most likely AGN sources 
found in machine learning through multi-wavelength analysis.
The companion work serves to cross check the results of the machine-learning 
method. Also, identification of individual sources could reveal hidden ones
that deserve full studies because of interesting
physical properties and processes.

In this paper, we report the identification of three new AGNs among 4FGL-DR4 
sources through our multi-wavelength analysis. They are 4FGL J0502.6+0036, 
4FGL  J1055.9+6507, and 4FGL J1708.2+5519.
This paper is organized as follows. Section \ref{sect:Data} describes 
the multi-wavelength data we used in the analysis. Section~\ref{sec:ar}
describes the analysis we conducted and provides the results.
We discuss the likely nature of the three sources in Section \ref{sect:dis}.

\section{Multi-wavelength Data}
\label{sect:Data}

\subsection{Optical and Mid-Infrared Data}

Optical imaging and optical light-curve data were obtained from 
the Panoramic Survey Telescope and Rapid Response System (Pan-STARRS;
\citealt{Chambers+16}) and the Zwicky Transient Facility (ZTF; 
\citealt{Bellm+19}) respectively. To obtain good-quality light curves,
we selected magnitude data points by requiring catflags = 0 and $chi < 4$;
the magnitudes are in the ZTF $g$- and $r$-band ($zg$ and $zr$ respectively).
In addition, 
the mid-infrared (MIR) light-curve data were obtained from the
NEOWISE Single-exposure Source Database (\citealt{Mainzer+14}), where
the bands are WISE w1 (3.4\,$\mu$m) and w2 (4.6\,$\mu$m).

The ZTF magnitudes are in the AB photometric system \citep{Bellm+19}, very
similar to those provided in Pan-STARRS \citep{ztf2pan}.
The MIR magnitudes are in the Vega system, and the conversion of them to 
fluxes can be calculated by 
using the given zero-magnitude flux density \citep{wri+10}. 

\subsection{X-ray Data}

For the fields of the three \fermi\ \gr\ sources, we searched for archival
X-ray data.
There were a few observations of the fields of 4FGL J0502.6+0036 and 4FGL J1055.9+6507
conducted with the X-Ray Telescope (XRT) onboard {\it the Neil Gehrels 
Swift Observatory (Swift)}. 
The observation information is summarized in Table~\ref{tab:obs}.
\begin{table}[!ht]
	\caption[]{Information for the {\it Swift}-XRT observations}
    \centering
    \begin{tabular}{cccc}
    \hline
        Date & Obsid & Exposure & $F^{unabs}_{0.3-10}/10^{-13}$ \\
        ~ & ~ & (Sec) & (erg cm$^{-2}$ s$^{-1}$) \\ \hline
        ~ & ~ & 4FGL J0502.6+0036 ~ ~ ~ ~ ~ & ~ \\ \hline
        2019-04-20 & 3107436001 & 1236 & ~ \\
        2024-02-15 & 3112553001 & 4323 & ~ \\
        2024-02-17 & 3112553003 & 5110 & ~ \\
        2024-02-19 & 3112553005 & 1799 & 1.4$^{+1.3}_{-0.6}$ \\
        2024-02-21 & 3112553007 & 576 & ~ \\
        2024-02-22 & 3112553009 & 287 & ~ \\
        2024-02-25 & 3112553011 & 1012 & ~ \\
        2024-02-27 & 3112553013 & 174 & ~ \\ \hline
        ~ & ~ & 4FGL  J1055.9+6507 ~ ~ ~ ~ ~ & ~ \\ \hline
        2024-03-06 & 3112558001 & 69 & ~ \\
        2024-03-15 & 3112558003 & 455 &  $\leq$10.6$^{*}$\\
        2024-04-04 & 3112558013 & 232 & ~ \\ \hline
    \end{tabular}
\tablecomments{0.86\textwidth}{$*$ indicates the 3$\sigma$ flux upper limit.}
	\label{tab:obs}
\end{table}

Using the online {\it Swift}-XRT data products generator tool\footnote{\url{https://www.swift.ac.uk/user_objects/}} (for details about the online tool,
see \citealt{Evans+07}; \citealt{Evans+09}; \citealt{Evans+20}), we ran source
detection in the XRT data. All the data for each of the two source fields
were combined, and only the counterpart to 4FGL J0502.6+0036 
(see below Section~\ref{subsec:src1}) was weakly detected. 
Its 0.3--10 keV spectrum was extracted. We fitted the spectrum with an absorbed 
power-law (PL) model in the XSPEC 12.12.1 using the C-Statistic method,
where the Galactic hydrogen column density $N_{\rm H}$ was fixed at 
$8.57 \times 10^{20}$\,cm$^{-2}$ \citep{HI4PICollaboration16}. 

For 4FGL J1055.9+6507, no counterpart was found (see below Section~\ref{subsec:src2}). 
Using the longest exposure (455 sec;
Table~\ref{tab:obs}), we obtained a 3$\sigma$ upper limit on the count rate
at the source position. To convert it to an upper limit on the 0.3--10\,keV 
flux,
we used the PIMMS tool\footnote{\url{https://cxc.harvard.edu/toolkit/pimms.jsp}}
by assuming a PL spectrum with a photon index of 2, where 
$N_{\rm H} = 2.25\times 10^{20}$\,cm$^{-2}$ \citep{HI4PICollaboration16}
towards the source direction was used.

\subsection{\fermi-LAT Data and Source Model}

The \fermi\ LAT data used were 0.1--500 GeV photon events 
(evclass=128 and evtype=3) from 
the updated Fermi Pass 8 database in a time range of from 2008-08-04 15:43:36 
(UTC) to 2023-11-01 09:31:26.8104 (UTC). The region of interest (RoI) for 
each target was set to be 20\degr $\times$ 20\degr\ centered at the position 
given in 4FGL-DR4. To avoid the contamination from the Earth limb, we 
excluded the events with zenith angles $>$ 90$^\circ$. We set the expression 
$DATA\_QUAL > 0 \&\& LAT\_CONFIG == 1$ to select good time-interval events.
In our analysis, the package Fermitools–2.2.0 and the instrumental response 
function P8R3\_SOURCE\_V3 were used.

The source models for the targets were generated based on 4FGL-DR4. 
The three targets were all modeled as a point source with a PL spectrum, 
$dN/dE = N{_0}(E/E{_0})^{-\Gamma}$, where $E{_0}$ were fixed at 1.74\,GeV, 
1.35\,GeV, and 1.4\,GeV, respectively, for 4FGL J0502.6+0036, 4FGL J1055.9+6507, and 4FGL J1708.2+5519. 
We adopted the same spectral models in our source models for them.
In addition, all 
sources in 4FGL-DR4 within 25\degr\ of a target were included.
We set the spectral indices and normalizations of 
the sources within 5\degr\ of a target as free parameters and fixed
the other parameters at the catalog values. We also included
the extragalactic diffuse emission and the Galactic diffuse emission components,
for which the spectral files iso\_P8R3\_SOURCE\_V3\_v1.txt and 
gll\_iem\_v07.fits were used respectively.
The normalizations of the two components were always set as the free
parameters in our analysis.

\begin{figure}[htbp]
  \begin{minipage}[t]{0.495\textwidth}
  \centering
   \includegraphics[width=73mm]{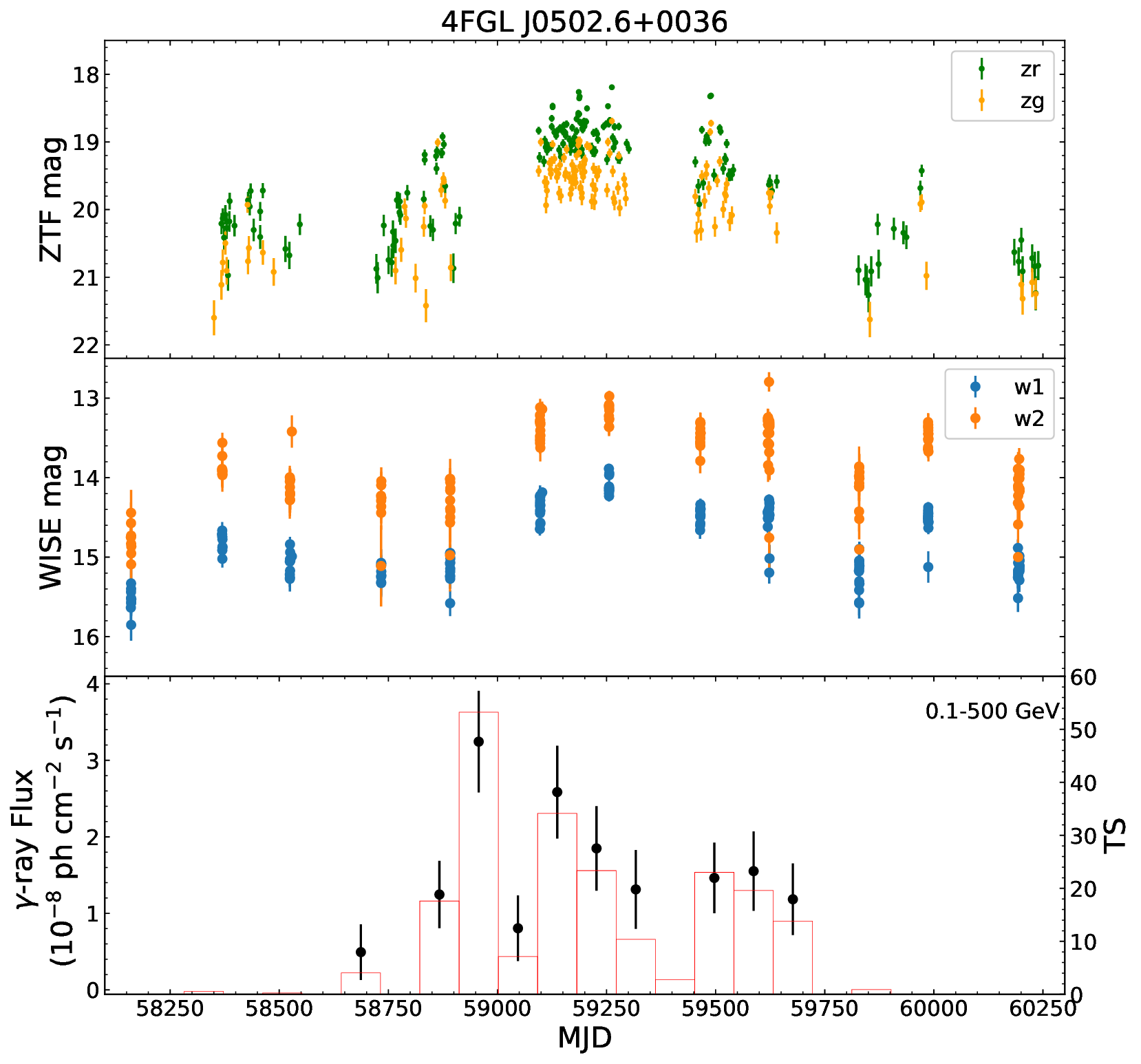}
  \end{minipage}%
  \begin{minipage}[t]{0.495\textwidth}
  \centering
   \includegraphics[width=75mm]{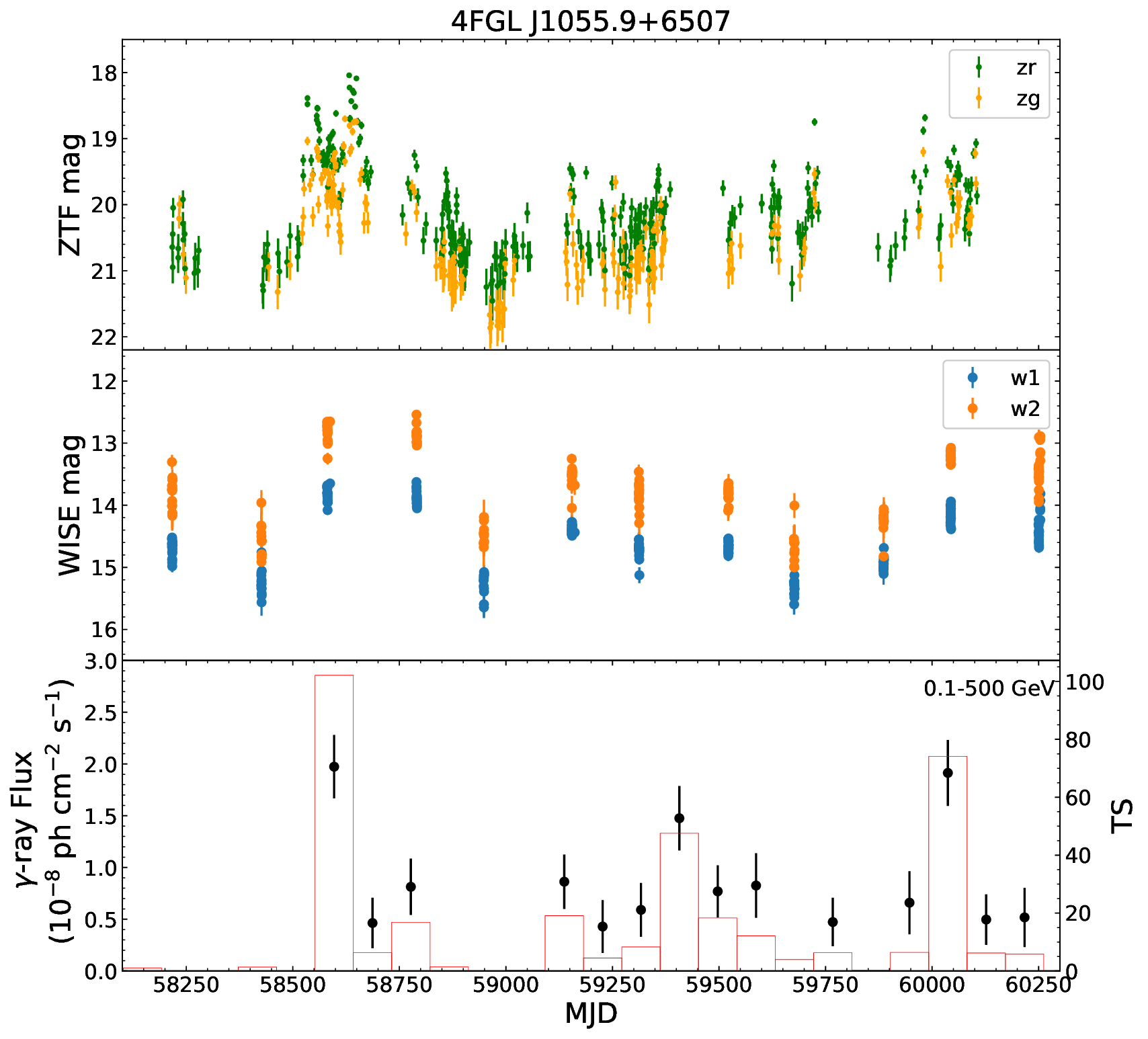}
  \end{minipage}%
  \qquad
   \begin{minipage}[t]{1\textwidth}
  \centering
   \includegraphics[width=75mm]{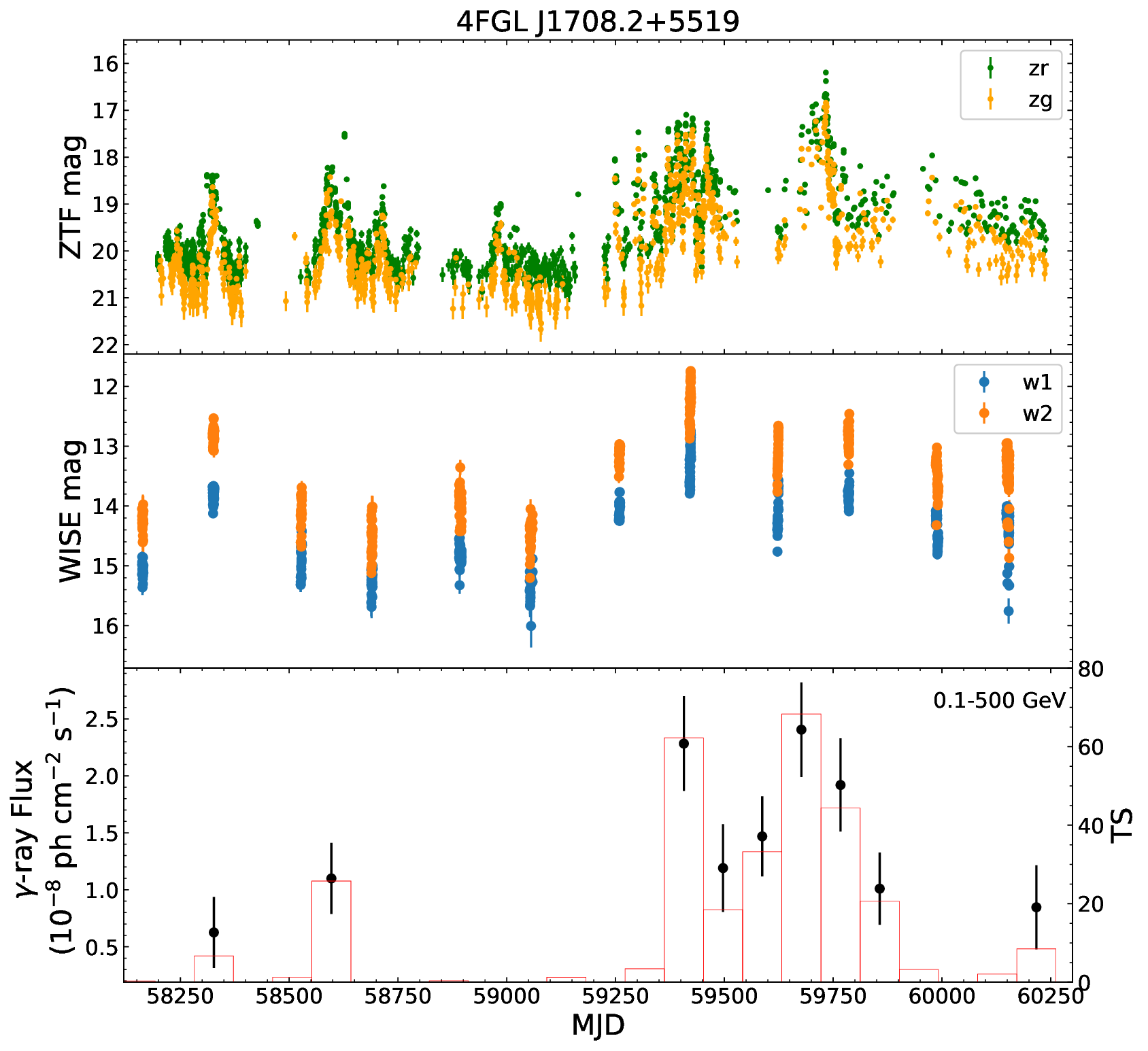}
  \end{minipage}%
\caption{\label{Fig1}{Multi-wavelength light curves for 4FGL J0502.6+0036, 4FGL J1055.9+6507,
	and 4FGL J1708.2+5519. The {\it top} and {\it middle} panels show the ZTF light 
	curves in $zg$ and $zr$ bands and WISE MIR light curves
	in w1 and w2 bands, respectively. The {\it bottom} panels show 
	90-day--binned light curves in 0.1--500\,GeV 
	(black dots), while the corresponding
	TS values are indicated by red bars. The data points with TS$<$4 are 
	not kept in the light curves.} }
\end{figure}

\section{Analysis and Results}
\label{sec:ar}

\subsection{4FGL J0502.6+0036}
\label{subsec:src1}

Using the source model described above, we performed the standard binned 
likelihood analysis to the data in 0.1--500\,GeV for 4FGL J0502.6+0036. 
The obtained best-fit $\Gamma = 2.13\pm$0.14, with
a test statistic (TS) value of $\simeq$54. The $\Gamma$ value is
consistent with that given in 4FGL-DR4, but the TS value is lower than
83 in 4FGL-DR4. The discrepancy in TS values is likely caused by our inclusion
of the latest data, during which the source was not detectable 
(see Figures~\ref{Fig1} \& \ref{fig:lc}).
With the best-fit model, we extracted the source's light curve
with 90-day a time bin.  In the extraction by
performing the maximum likelihood analysis to the data of each time bin,
only the normalization parameters of
the sources within 5\degr\ of the target were set free and the other 
parameters were fixed at the best-fit values obtained in the analysis of 
the whole data. The whole light curve is shown in Figure~\ref{fig:lc}.
As can be seen, there was a flare during the approximate time period of
MJD~58600--59750 (see Figure \ref{Fig1}). We also calculated
the variability index TS$_{var}$ \citep{Nolan+12} of the light curve, 
and obtained
TS$_{var}$ $\approx$179.3. The index value indicates variability of
the source at a 7.4$\sigma$ significance level (as TS$_{var}$ is considered 
to be distributed as $\chi^2$, with 61 degrees of freedom). 

\begin{figure}[htbp]
  \begin{minipage}[t]{0.5\linewidth}
  \centering
   \includegraphics[width=75mm]{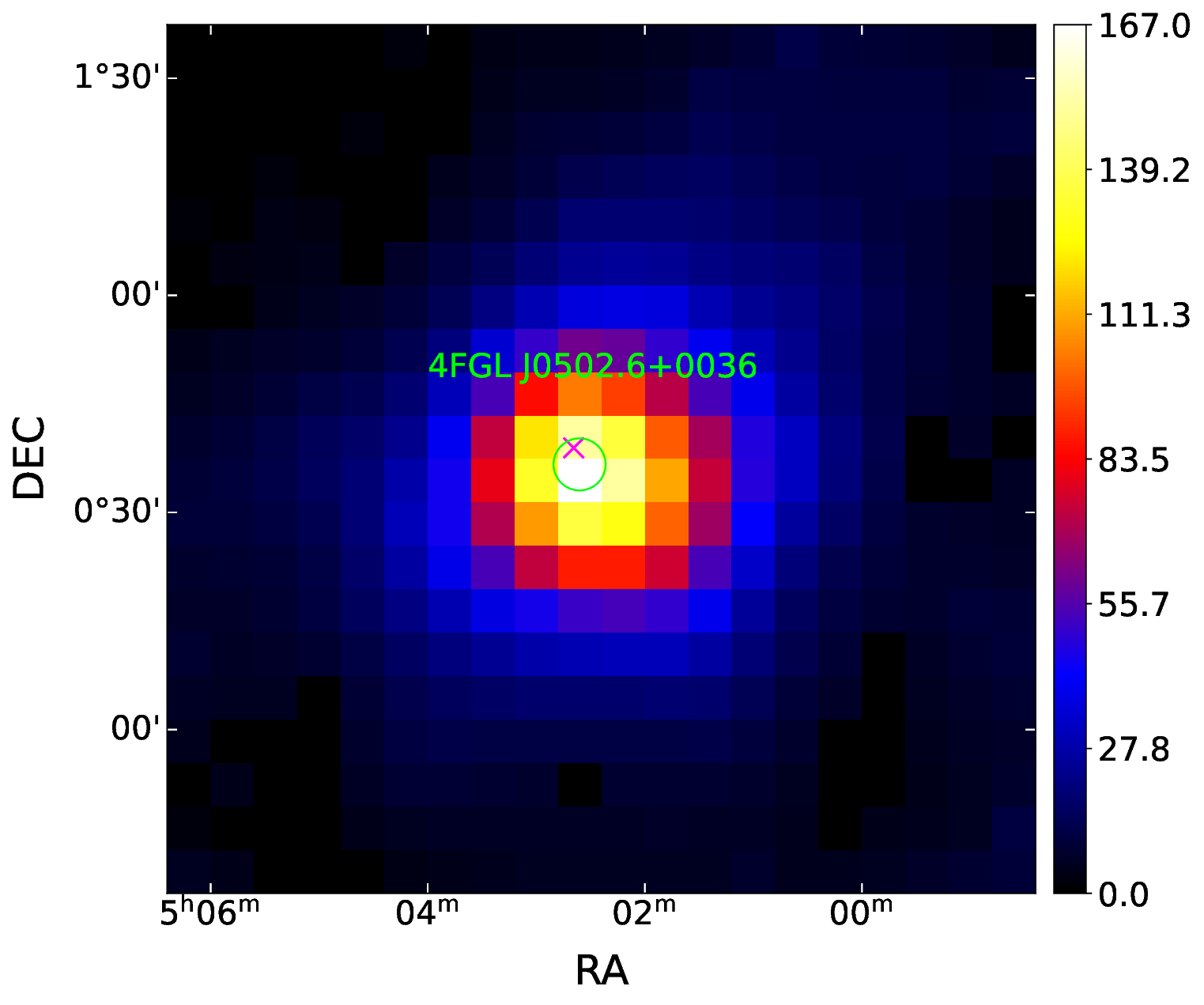}
  \end{minipage}%
  \begin{minipage}[t]{0.5\textwidth}
  \centering
   \includegraphics[width=69mm]{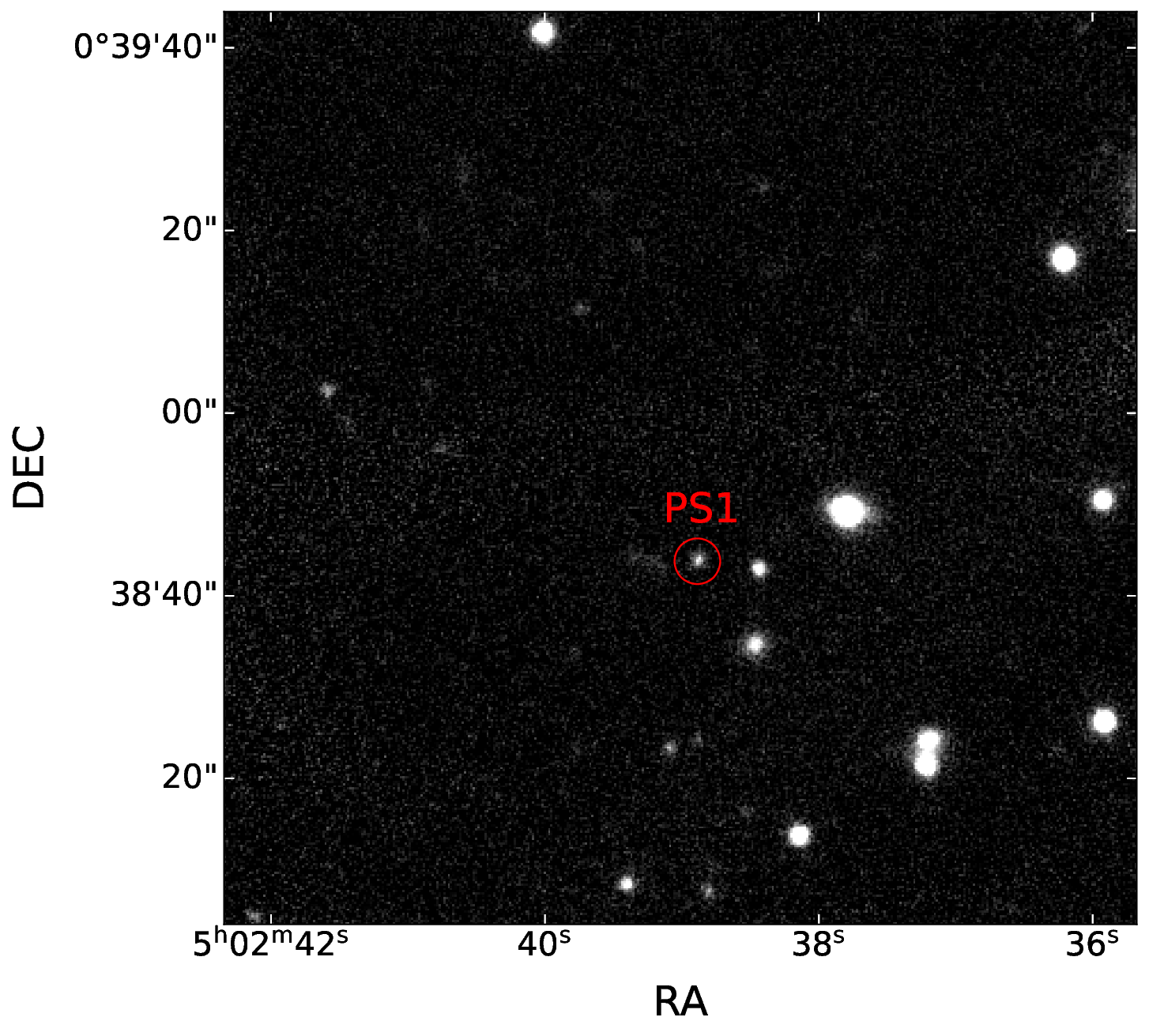}
  \end{minipage}%
	\caption{\label{Fig2}{{\it Left}: 0.1--500 GeV $2\degr \times 2\degr$ 
	TS map centered at 4FGL J0502.6+0036 during MJD~58600--59750, with its
	2$\sigma$ positional error circle marked as a green circle. 
	The purple cross marks the location of the radio source BWE 0500+0034. 
	{\it Right}: Pan-STARRS $i$-band 
	$100^{\prime \prime} \times 100^{\prime \prime}$ image centered at 
	BWE~0500+0034. The optical source PS1 is indicated by a red circle.} }
\end{figure}

To connect 4FGL J0502.6+0036 with its multi-wavelength counterparts, we ran 
{\tt gtfindsrc} in Fermitools to determine the source's position.
We obtained R.A.$=$ 75\degr.65, Decl.$=$+0\degr.61 (equinox J2000.0), 
with a 2$\sigma$ uncertainty of 0\degr.06. This position is consistent
with that given in 4FGL-DR4 (see Table~\ref{tab:loc}). 
Within the error circle of the position,
we found only one radio source, BWE 0500+0034, in the SIMBAD database. 
We used the tool {\tt gtsrcid} in Fermitools \citep{Abdo+10} to 
estimate the probability of 4FGL J0502.6+0036 in association with BWE 0500+0034.
The calculated probability was 97.8\%, suggesting a highly likely association 
between them, given that {\it Fermi} LAT used a threshold of 80\% to determine 
associations for point sources. The 0.1--500\,GeV TS map of the source field, 
centered at 4FGL J0502.6+0036, 
during the flare time period (MJD~58600--59750) was calculated. The TS map,
shown in the left panel of Figure \ref{Fig2}, helps indicate the positional
coincidence between 4FGL J0502.6+0036 and BWE 0500+0034. The latter has a 95\%-confidence
radius of $\sim$ 50$^{\prime \prime}$ (\citealt{Becker+1991}).

Using the Pan-STARRS $100^{\prime \prime} \times 100^{\prime \prime}$ image 
centered at BWE 0500+0034 (right panel of Figure~\ref{Fig2}), we examined
optical sources in the field for the potential counterpart. 
There are more than ten sources, and we obtained their ZTF and WISE MIR 
light curves to check their flux variations.
The optical source closest to the center of BWE 0500+0034, called PS1 in
this work (Figure~\ref{Fig2}), showed concurrent brightening activity during
the flaring time period of 4FGL J0502.6+0036 (Figure~\ref{Fig1}).
Because of the temporal and spatial coincidence, BWE 0500+0034 and PS1 
are thus the likely radio and optical counterpart.

\begin{table}[!ht]
	\caption[]{Comparison of source information in this work and in 4FGL-DR4}
    \centering
    \begin{tabular}{ccccccc}
    \hline
        Name & Reference & R.A. & Decl. & $R_{95}$ & $\Gamma_\gamma$ & TS\\
        ~ & ~ & deg & deg & deg \\\hline
        4FGL J0502.6+0036 & 4FGL-DR4 & 75.65 & +0.61 & 0.05 & 2.24$\pm$0.11 & 83\\ 
        4FGL J0502.6+0036  & This work & 75.65 & +0.61 & 0.06 & 2.13$\pm$0.14 & 54\\
        BWE 0500+0034 & \citealt{Becker+1991} & 75.66 & +0.65 &  \\\hline
        4FGL J1055.9+6507 & 4FGL-DR4 & 163.98 & +65.13 & 0.05 & 2.32$\pm$0.08 & 198\\
        4FGL J1055.9+6507 & This work & 163.92 & +65.14 & 0.04 & 2.25$\pm$0.08 & 168\\
        NVSS J105533+650956 & \citealt{Condon+1998} & 163.89 & +65.17 & \\\hline
        4FGL J1708.2+5519 & 4FGL-DR4 & 257.05 & +55.32 & 0.09 & 2.38$\pm$0.09 & 167\\
        4FGL J1708.2+5519 & This work & 257.05 & +55.32 & 0.04 & 2.41$\pm$0.11 & 171\\
        FIRST J170802.8+551920 & \citealt{Helfand+15} & 257.01 & +55.32 &  \\\hline
    \end{tabular}
\tablecomments{0.86\textwidth}{$R_{95}$ is the error radius at a 95\% confidence level; the value of 4FGL-DR4 is the average one estimated from values of the semi-major axis and semi-minor axis of the error ellipse at a 95\% confidence level.}
	\label{tab:loc}
\end{table}

\subsection{4FGL J1055.9+6507}
\label{subsec:src2}

By performing the standard binned likelihood analysis to the data in 
0.1--500\,GeV for 4FGL J1055.9+6507, we obtained the best-fit parameters, in which
$\Gamma= 2.25\pm$0.08, with a TS value of $\simeq$ 168. The values
are in agreement with 
those given in 4FGL-DR4. Using the same setting as those for 4FGL J0502.6+0036, 
we extracted its 90-day binned light curve (see Figure~\ref{fig:lc}). 
The light curve also shows flare-like events, which mostly occurred 
during MJD~58500--60250 in the recent years (Figure~\ref{Fig1}). 
The variability index TS$_{var}$ was found
to be $\approx$260.2, which indicates source variability at a 10.6$\sigma$ 
significance level. Thus, the source was also significantly variable.

\begin{figure}[htbp]
  \begin{minipage}[t]{0.495\linewidth}
  \centering
   \includegraphics[width=75mm]{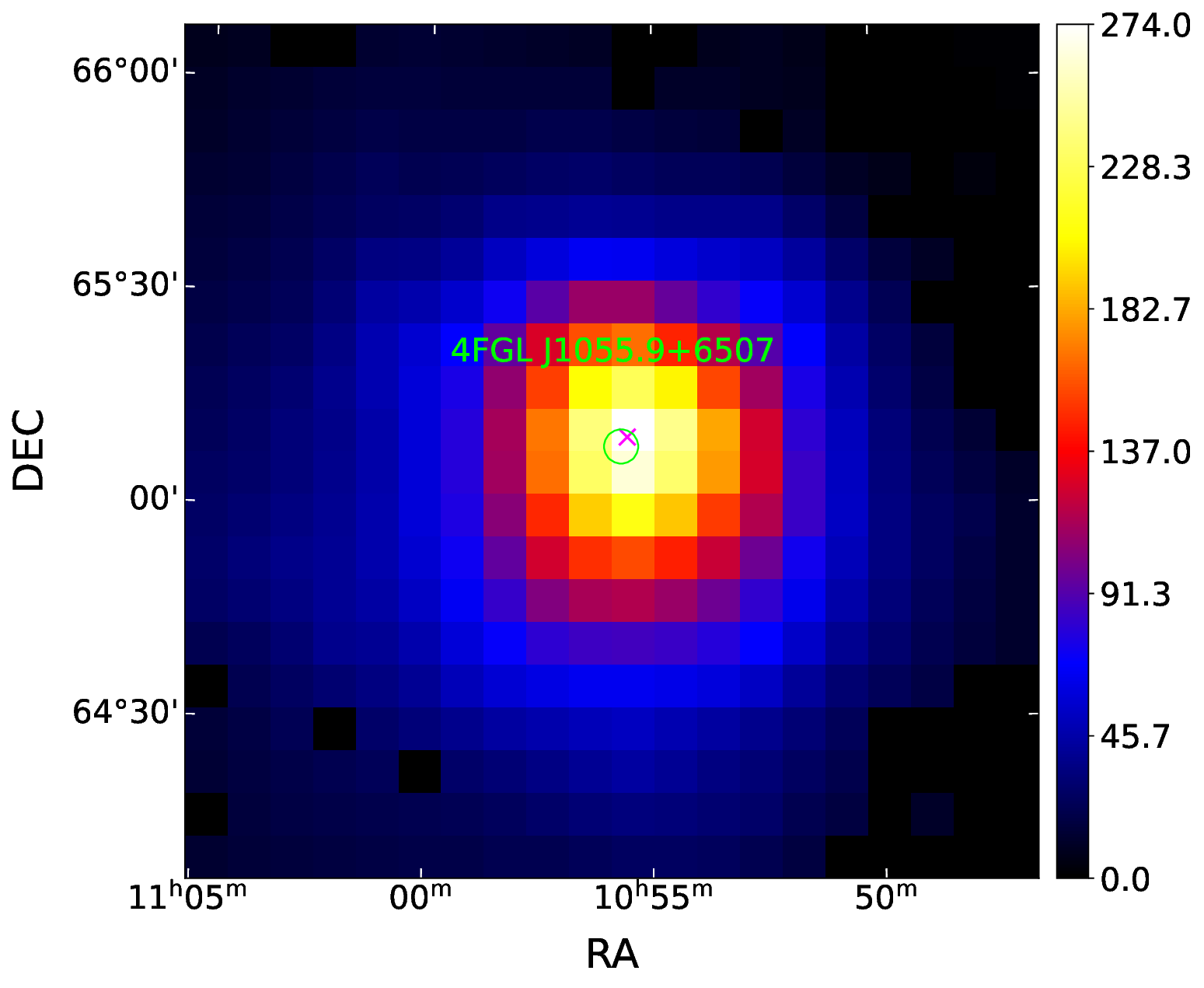}
  \end{minipage}%
  \begin{minipage}[t]{0.495\textwidth}
  \centering
   \includegraphics[width=69mm]{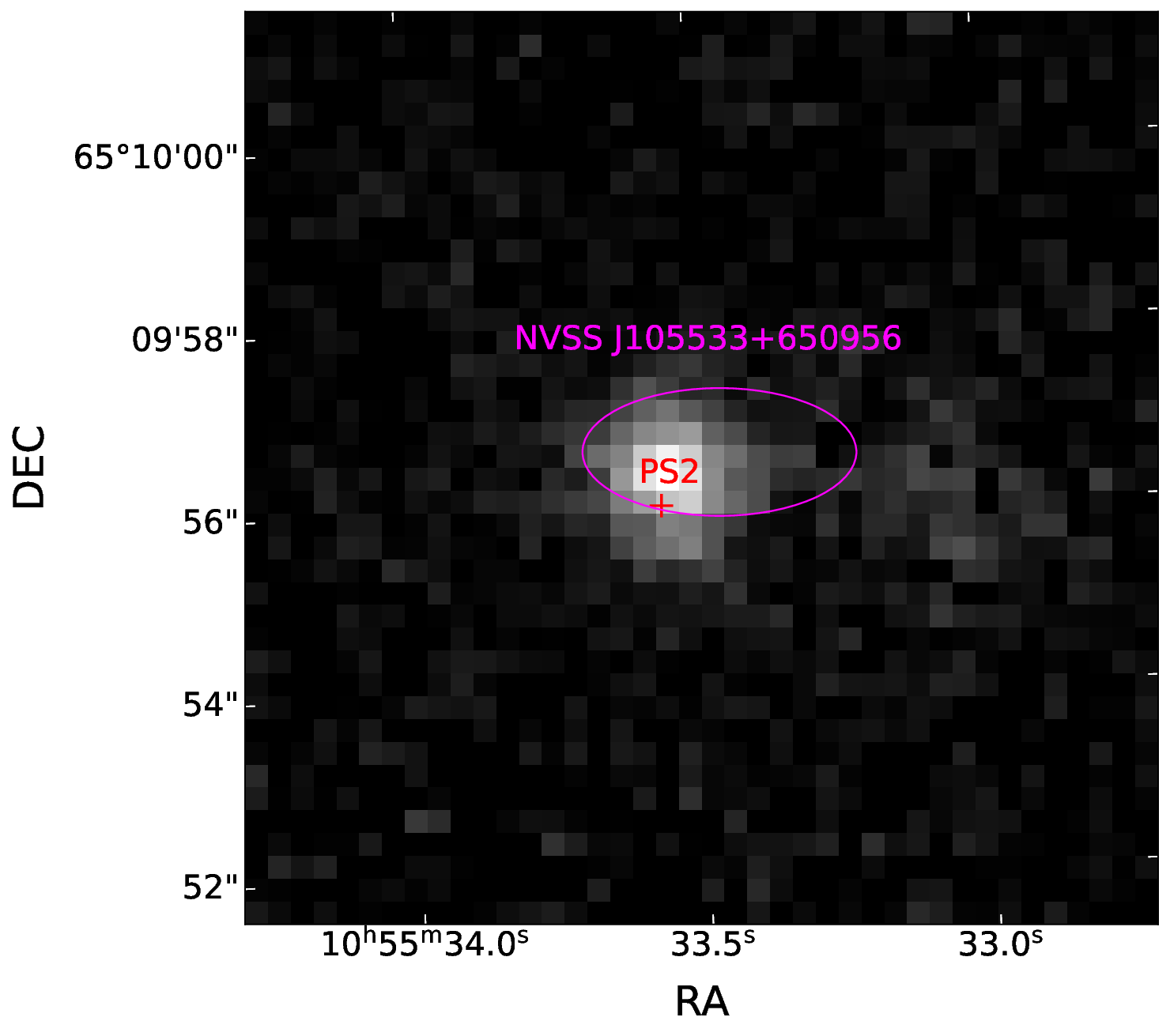}
  \end{minipage}%
	\caption{\label{Fig3}{{\it Left}: 0.1--500 GeV $2\degr \times 2\degr$ 
	TS map centered at 4FGL J1055.9+6507, with its 2$\sigma$ positional error circle
marked as a green circle. The purple cross is the location of the radio source 
	NVSS~J105533+650956. {\it Right}: Pan-STARRS $i$-band 
	$10^{\prime \prime} \times 10^{\prime \prime}$ image centered at 
	NVSS~J105533+650956. The positional error region of 
	NVSS J105533+650956 is indicated by a purple ellipse, and
	the optical source PS2 is marked with a red cross. There is a 0$''$.3
	offset between the peak of the optical source and the position
	of PS2 given in Pan-STARRS.} }
\end{figure}

Running {\tt gtfindsrc} to check the source's position, we obtained 
R.A.$=$163\degr.92, Decl. $=$ +65\degr.14 (equinox J2000.0), with 
a 2$\sigma$ uncertainty of 0\degr.04 (see Table \ref{tab:loc}). Further checking radio sources
as the possible candidate counterpart, there is only one found,
given in the NRAO VLA Sky Survey (NVSS) catalog (\citealt{Condon+1998}) as
NVSS~J105533+650956. Running {\tt gtsrcid}, we obtained an association
probability of 98.8\% between 4FGL J1055.9+6507 and NVSS J105533+650956, also 
indicating their association is a highly likely case.
We also calculated the TS map in 0.1--500\,GeV during the flaring time period
for the source field (left panel of Figure~\ref{Fig3}). The TS map illustrates
the positional matches between 4FGL J1055.9+6507 and the radio source.
The Pan-STARRS $10^{\prime \prime} \times 10^{\prime \prime}$ image 
(Figure~\ref{Fig3}), in
$i$-band centered at NVSS J105533+650956, was obtained to search for the 
possible optical counterpart. There is one source, named PS2, 
in the error region of NVSS~J105533+650956. We note that PS2, with its
position given in Pan-STARRS as the average one from multiple exposures in
multiple bands, is 0$''$.3 off the peak of the optical source 
in the $i$-band image (Figure~\ref{Fig3}). This offset could reflect
the systematic positional uncertainty in astrometry of Pan-STARRS
(a nearly the same offset was found in the next case 4FGL J1708.2+5519; See 
Section~\ref{subsec:j17}).
The optical source's ZTF and WISE light curves were obtained
and are shown in Figure~\ref{Fig1}. The multi-wavelength light curves
show similar variability behavior as the \gr\ one. When the optical
and MIR magnitudes were at the brightest during MJD~58500--58800, 
with the optical showing strong variation activity, 
the \gr\ flux 
(and TS value) was also at the highest. Given both the spatial and
temporal coincidence, NVSS J105533+650956 and PS2 are highly likely the radio 
and optical counterparts to 4FGL J1055.9+6507.

\subsection{4FGL J1708.2+5519}
\label{subsec:j17}

We performed the standard binned likelihood analysis to the data 
in 0.1--500\,GeV for 4FGL J1708.2+5519, and the obtained best-fit 
$\Gamma= 2.41\pm$0.11, with a TS value of $\simeq$ 171, in agreement 
with those given in 4FGL-DR4. Using the same setting as that for
the above two sources, its 90-day binned light curve was extracted 
(see Figure~\ref{fig:lc}). The light curve shows flare events, with a
dominant one occurring approximately during MJD~59250--60000.
The calculated TS$_{var}\approx$196.3, indicating \gr\ flux 
variations of the source at an 8.1$\sigma$ significance level.

\begin{figure}[htbp]
  \begin{minipage}[t]{0.495\linewidth}
  \centering
   \includegraphics[width=75mm]{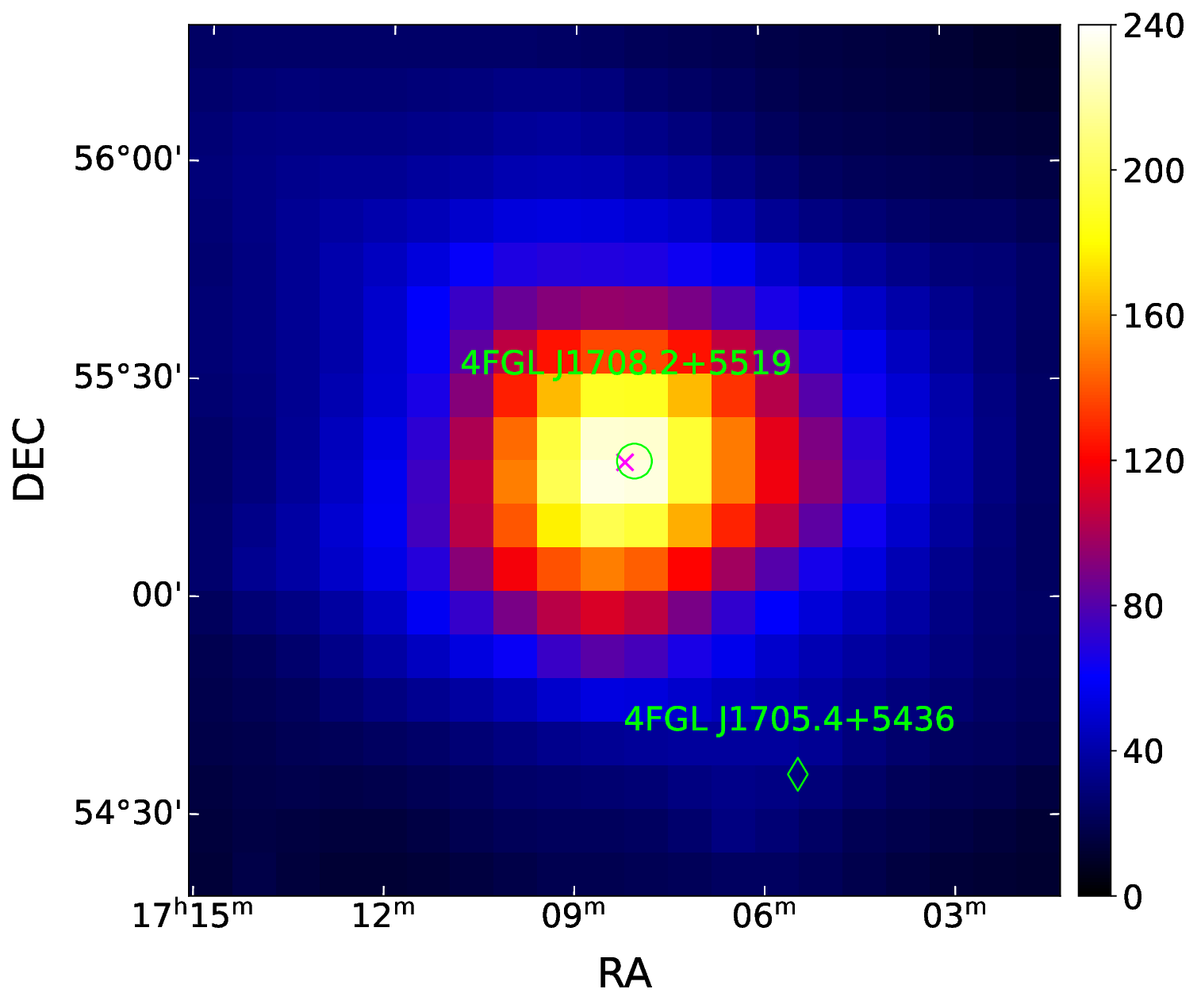}
  \end{minipage}
  \begin{minipage}[t]{0.495\linewidth}
  \centering
   \includegraphics[width=72mm]{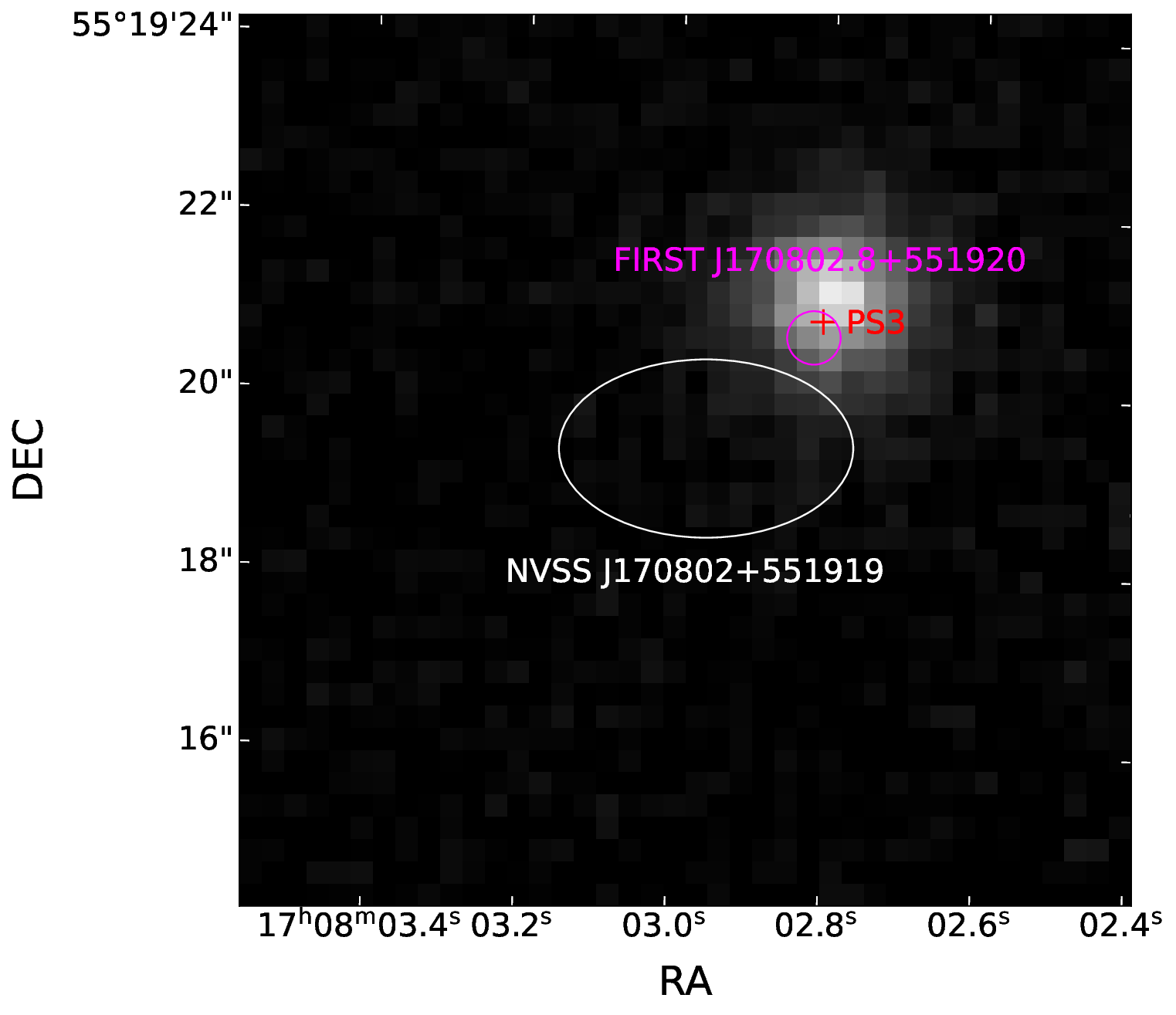}
  \end{minipage}
	\caption{\label{Fig4}{{\it left}: 0.1--500 GeV $2\degr \times 2\degr$ 
	TS map centered at 4FGL J1708.2+5519 during the flaring time period
	of MJD~59250--60000. The source's 2$\sigma$ positional error circle
	is marked as a green circle, and the purple cross marks the location 
	of the radio source FIRST~J170802.8+551920. {\it Right}: Pan-STARRS 
	$i$-band $10^{\prime \prime} \times 10^{\prime \prime}$ image 
	centered at NVSS J105533+650956, whose error region is
	marked by the white solid ellipse. The positional error circle 
	of FIRST J170802.8+551920 is marked by a purple circle, which contains
	the optical source PS3 (the red cross). There is a 0$''$.3 offset 
	between the peak of the optical source and the position of PS3 given
	in Pan-STARRS.} }
\end{figure}

From running {\tt gtfindsrc}, we obtained R.A.$=$ 257\degr.05, 
Decl.$=$ +55\degr.32 (equinox J2000.0), with a 2$\sigma$ uncertainty 
of 0\degr.04 (see Table \ref{tab:loc}). 
Within the error circle of the position, there is one radio source,
NVSS J170802+551919, given in the SIMBAD database. The calculated
TS map in 0.1--500\,GeV during the dominant-flare time period
(MJD~59250--60000) shows
the positional match between 4FGL J1708.2+5519 and the radio source
(left panel of Figure \ref{Fig4}). There is another radio source
from the Faint Images of the Radio Sky at Twenty-cm (FIRST) survey,
FIRST J170802.8+551920, which is slightly away. The FIRST survey typically has
a positional uncertainty of $0^{\prime \prime}.3$ (\citealt{Helfand+15}). 
We note that these two radio sources were matched as one in, for example,
\citet{Flesch+24}. A 99.2\% probability for the association between 
4FGL J1708.2+5519 and FIRST J170802.8+551920 was estimated with {\tt gtsrcid}, 
indicating their association is a highly likely case. The Pan-STARRS 
$10^{\prime \prime} \times 10^{\prime \prime}$ image in $i$-band centered 
at NVSS J170802+551919 is shown in the right panel of Figure~\ref{Fig4}. 
There is one optical source, named PS3 in this work, positionally matches
FIRST J170802.8+551920. Here again, the position of PS3 given in
Pan-STARRS is 0$''$.3 away from the peak of optical source in
the $i$-band image.  The ZTF and WISE light curves of PS3 
(Figure \ref{Fig1}) show concurrent variations with the \gr\ flaring events, 
especially the dominant one in MJD~59250--60000.
Therefore given the closeness
of the radio and optical sources and the similarity in temporal variations
at multi-wavelengths, it is highly likely that the two radio sources are 
the same one and they and PS3 are the radio and optical counterparts of 4FGL J1708.2+5519.

\subsection{Broadband spectral energy distributions of the three sources}

Given the identification established above, we built the broadband spectral 
energy distributions (SEDs) for the three sources. First
we extracted the $\gamma$-ray spectrum for for each of them. 
Using the best-fit models obtained above from the likelihood analysis, 
we performed the maximum likelihood analysis to the data in 10 evenly 
divided energy bins in logarithm from 0.1 to 500 GeV. In this analysis,
the spectral normalizations of the sources in a source model within 5\degr\ 
of each target source were set free, and all other spectral parameters 
of the sources were fixed at the best-fit values obtained in
the likelihood analysis of the whole data. For the extracted $\gamma$-ray 
spectra, we only kept the data points with TS $\geq$ 4.
\begin{figure}[ht]
  \begin{minipage}[t]{0.495\linewidth}
  \centering
   \includegraphics[width=75mm]{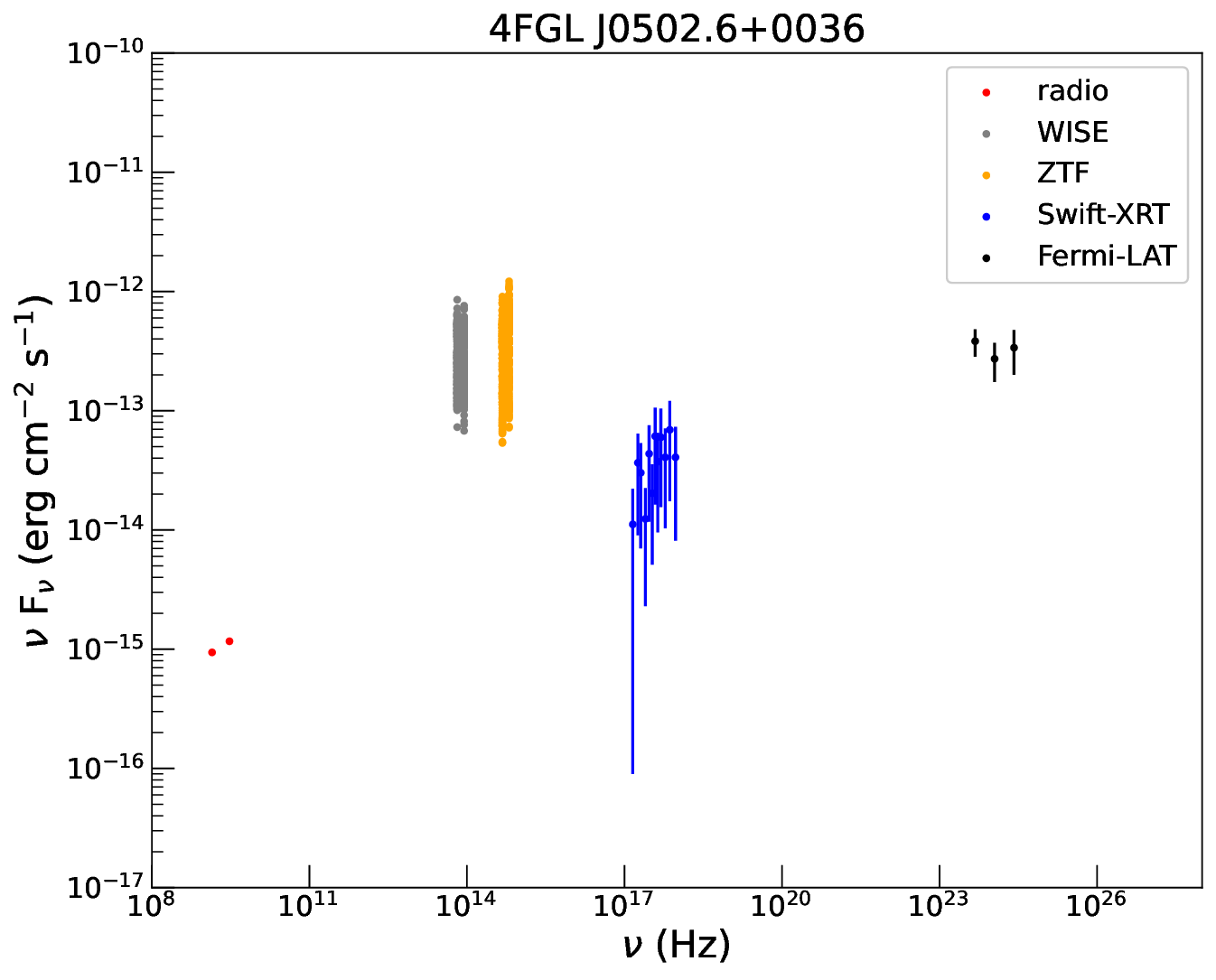}
  \end{minipage}%
  \begin{minipage}[t]{0.495\textwidth}
  \centering
   \includegraphics[width=75mm]{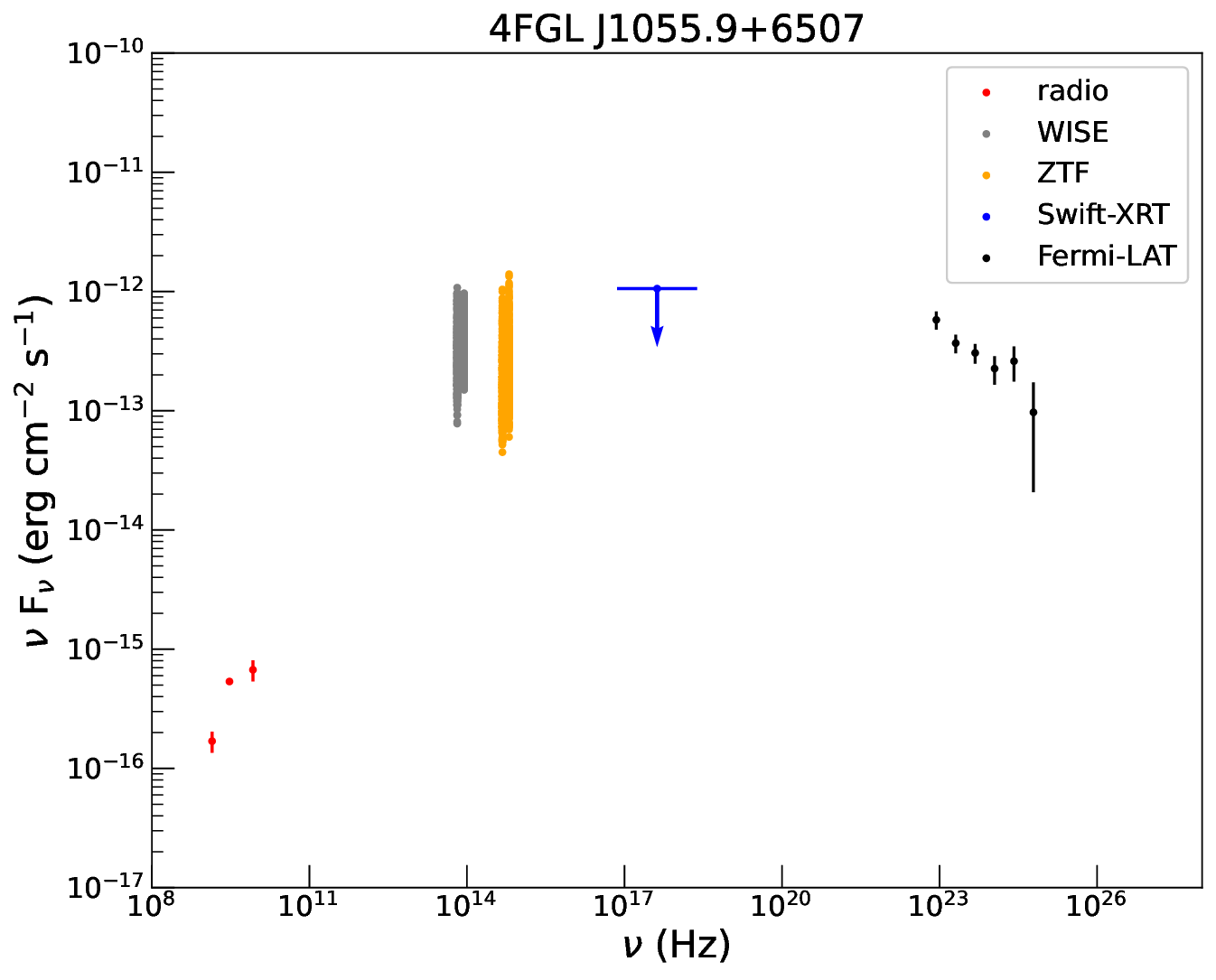}
  \end{minipage}%
  \qquad
   \begin{minipage}[t]{1\textwidth}
  \centering
   \includegraphics[width=75mm]{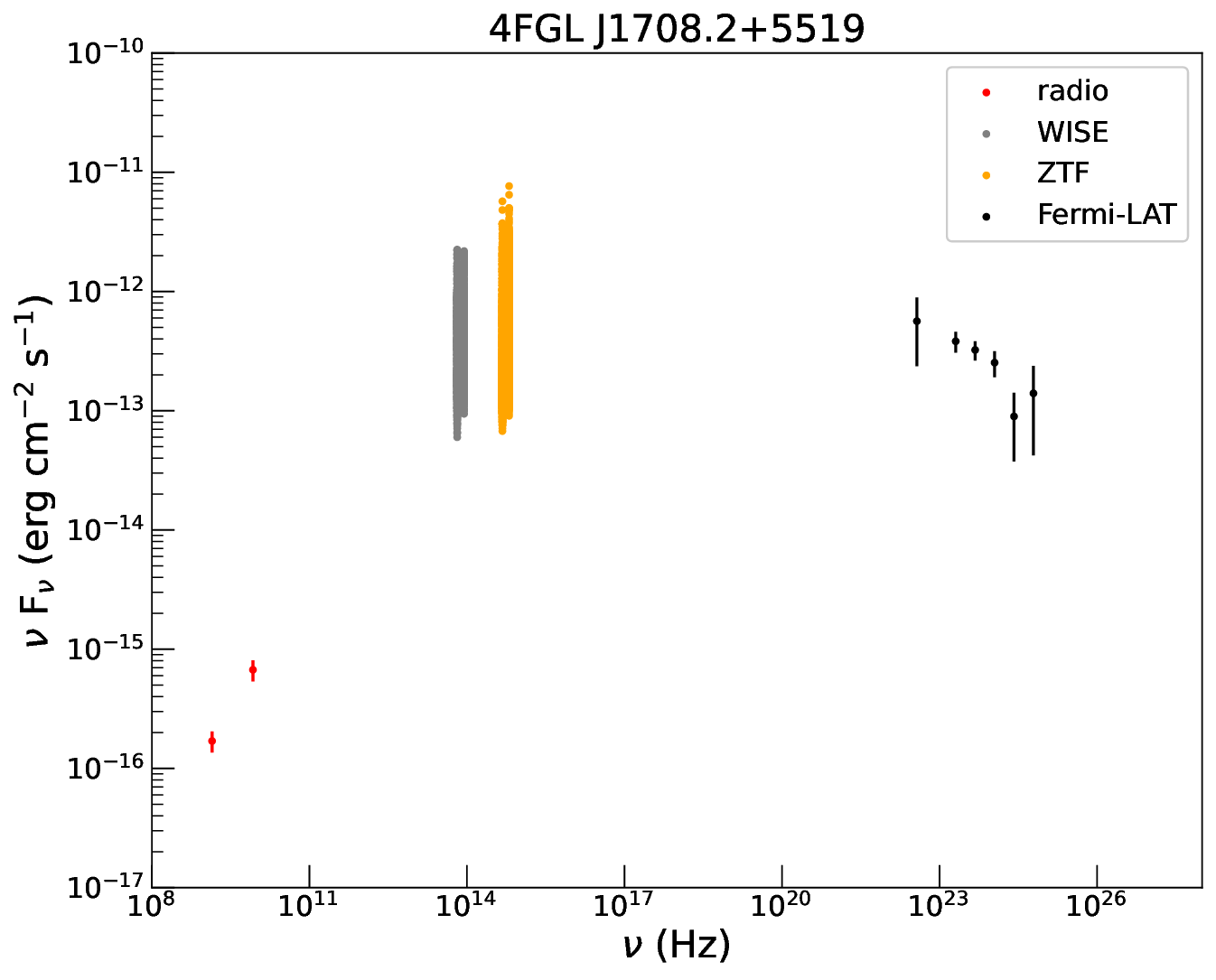}
  \end{minipage}%
\caption{\label{Fig5}{Broadband SEDs of 4FGL J0502.6+0036, 4FGL J1055.9+6507, and 4FGL J1708.2+5519. 
	Archival radio flux measurements are red dots (from 
	\citealt{Condon+1998}; \citealt{Helfand+15}; \citealt{Myers+03}; 
	\citealt{Gordon+21}), and MIR w1 and w2 and ZTF $zr$ and $zg$ fluxes 
	(used in this work) are grey and yellow dots respectively. 
	The {\it Swift}-XRT X-ray spectrum of 4FGL J0502.6+0036 is plotted as
	the blue dots (top-left panel), and the X-ray flux upper limit 
	for 4FGL J1055.9+6507 is indicated by a blue arrow (top-right panel).
	The $\gamma$-ray spectra we obtained for the three sources
	are plotted as black dots in each panel.}}
\end{figure}

The broadband SEDs of blazars often show two emission humps, a low-energy hump
peaking at radio to ultraviolet/X-ray wavelengths and the other one at
hard X-ray to \gr\ energies. In the leptonic model that is widely considered, 
the low-energy hump is produced from synchrotron radiation of
relativistic electrons in jets while the high-energy one is generated by 
inverse Compton (IC) scattering of low-energy photons by the same population
of the electrons (e.g., \citealt{Sikora+1994}; \citealt{Bloom+1996}).
Given the X-ray detection of 4FGL J0502.6+0036, its SED appears to be in such a two-hump
shape. We did not attempt to fit the SED with a typical leptonic model
for blazars, since the data were not simultaneous and had strong variations.
For 4FGL J1055.9+6507 and 4FGL J1708.2+5519, their SEDs share the similarity with that of 4FGL J0502.6+0036,
although X-ray detection would help strengthen it. We checked the eROSITA
X-ray data \citep{eRo+21}, but unfortunately both 4FGL J1055.9+6507 
and 4FGL J1708.2+5519 are not in the sky region of the recently released survey data.

\section{Discussion}
\label{sect:dis}

\begin{figure}[h]
  \begin{minipage}[t]{0.495\linewidth}
  \centering
   \includegraphics[width=75mm]{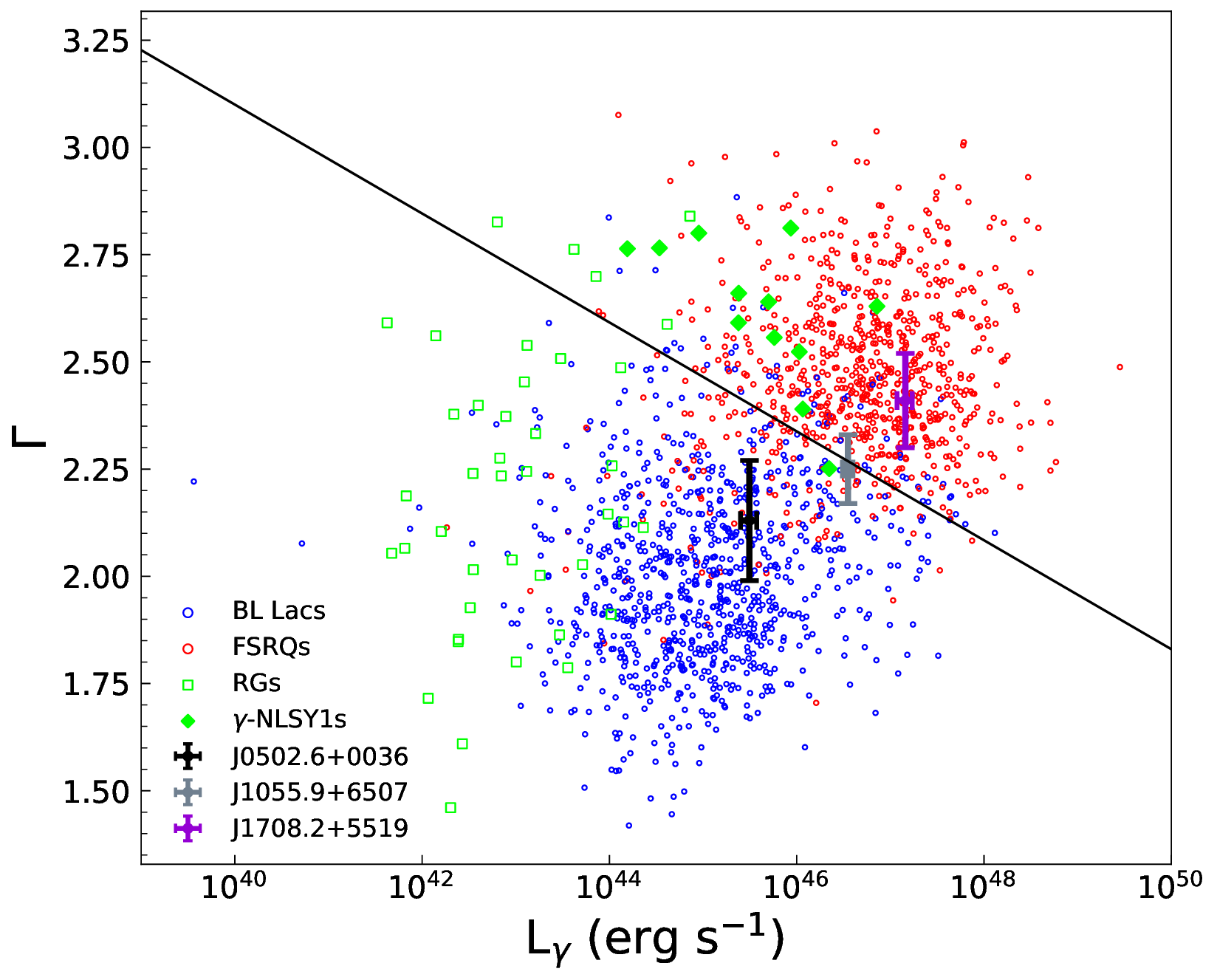}
  \end{minipage}%
  \begin{minipage}[t]{0.495\textwidth}
  \centering
   \includegraphics[width=76mm]{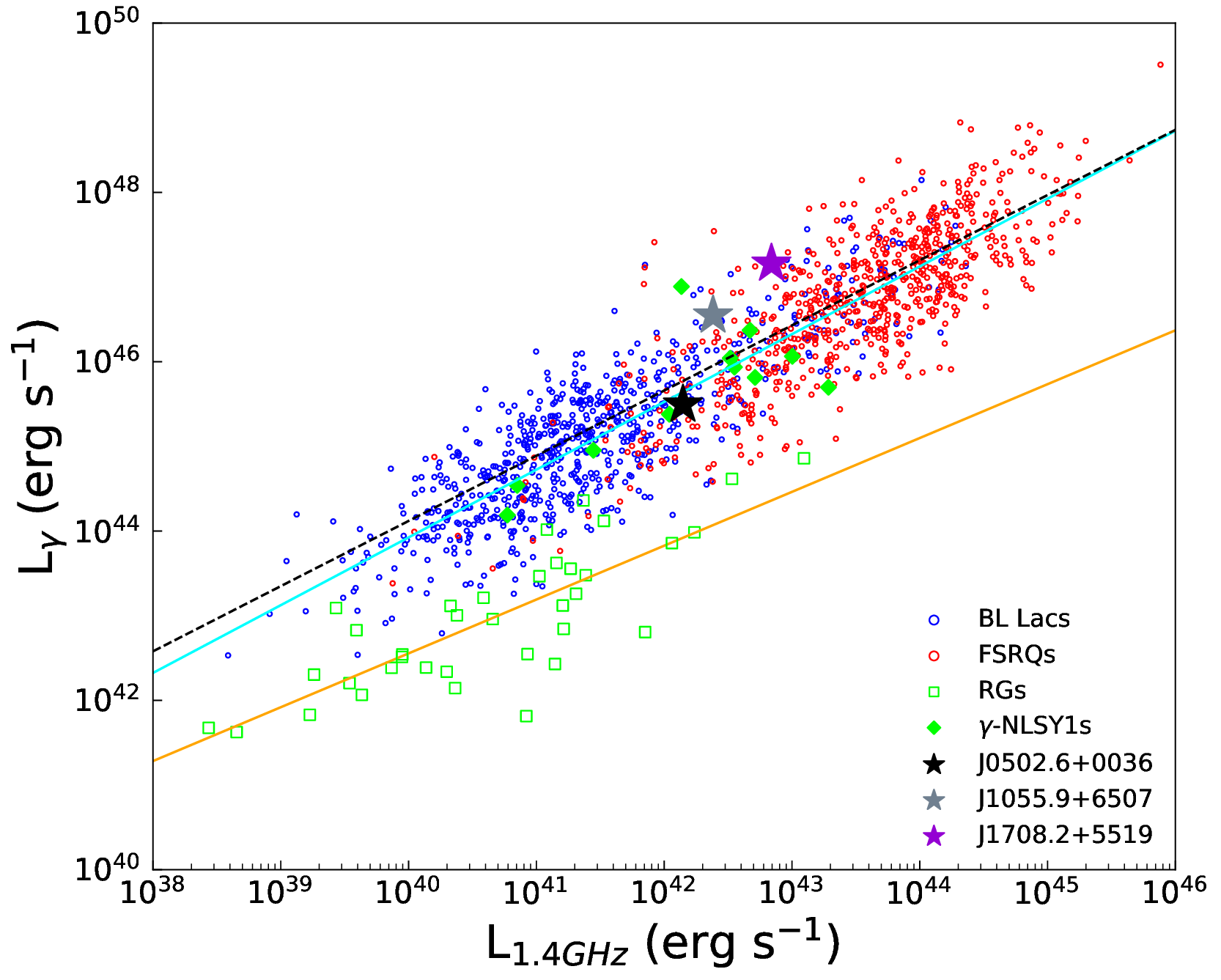}
  \end{minipage}%
\caption{{\it Left}: $\gamma$-ray photon indices versus $\gamma$-ray 
luminosities of 
different sub-types of AGNs, drawn from the information provided in 4LAC-DR3 
\citep{Ajello+22}. The three sources in this work are marked.
FSRQs and BL Lacs are possibly separated by a black line 
($\Gamma = -0.127 log L_{\gamma} + 8.18$; \citealt{Chen18}). 
{\it Right}: $\gamma$-ray luminosities versus 1.4\,GHz radio 
luminosities of $\gamma$-ray AGNs. The radio measurements 
	at 1.4 GHz were obtained from the NED. The black dashed line is given 
	by \citet{Wu+24} for blazars 
	($\log L_{\gamma} = 0.77 \log L_{\rm 1.4GHz} + 13.32$). 
	The cyan and orange solid lines are the relationships 
	for blazars and RGs, respectively, we obtained in this work.}
\label{Fig6}
\end{figure}

Among the AGNs detected by \fermi-LAT, blazars are the dominant ones
(e.g., \citealt{Ballet+23}). In addition, there are also so-called Narrow Line 
Seyfert 1 (NLSy1) galaxies and RGs. The former are defined by their optical 
narrow Balmer (the full-width at half maximum of H$_\beta$ 
$<$2000\,km\,s$^{-1}$), weak [O III], and strong Fe II emission lines
(\citealt{Osterbrock+1985}; \citealt{Goodrich+1989}). They are considered 
to be able
to host relativistic jets similar to blazars (e.g., \citealt{Abdo+2009}; \citealt{Paliya+14} and 
reference therein) and thus exhibit multi-wavelength variability as well
(e.g., \citealt{Abdo+2009}; \citealt{Paliya+13}; \citealt{Paliya+14}; \citealt{Paliya+16}). 
At the \gr\ band, they show soft emission ($\Gamma > 2$), similar to FSRQs, but
have \gr\ luminosities $L_{\gamma}$ smaller than those of powerful FSRQs
(\citealt{Paliya+18}; see also Figure~\ref{Fig6}). Studies of
their broadband SEDs also show that they more closely resemble FSRQs than 
BL Lacs (e.g., \citealt{Zhang+13,Sun+15}). The latter, RGs,
are considered as blazars with a misaligned jet. Most of RGs have low redshifts;
for example, the recorded most distance RG has a redshift of 0.22 in
the \fermi-LAT sample \citep{Ajello+22}. As a result, their \gr\ luminosities 
are generally lower
than those of blazars (Figure~\ref{Fig6}). We note that \citet{Paliya+23} 
recently have identified a distant $\gamma$-ray RG, TXS 1433+205, 
at $z =$ 0.748, but it can be considered as an extreme case.

We checked the radio images for the three sources, and they all only 
show a core, without any structures seen. In the more recent radio data 
from the Australian Square 
Kilometre Array Pathfinder (ASKAP; \citealt{Johnston+07,Johnston+08,Hotan+21}),
4FGL J0502.6+0036 is within the survey's field. There were six 
flux-density measurements at 887.5\,MHz and one at 1367.5\,MHz. The lowest 
and highest flux densities at 887.5\,MHz were 52.7$\pm$0.4 mJy 
and 74.4$\pm$0.5 mJy, respectively; the flux density at 1367.5\,MHz
was 57.4$\pm$0.4 mJy, while the NVSS's flux density was 67.2$\pm$0.9 mJy 
at 1.4\,GHz.  These measurements show that the source had significant flux
variations at radio frequencies as well.

By the large fraction of blazars in the \fermi-LAT sample, the three sources
in this work are likely blazars. Their multi-wavelength variabilities and
SEDs all fit in this identification.  In \citet{Flesch+24}, the redshifts
of the optical counterparts to
4FGL J0502.6+0036, 4FGL J1055.9+6507, and 4FGL J1708.2+5519 are given to 
be 0.6, 1.5, and 2.3, respectively, derived from photometric measurements.
With the redshifts assumed,
their locations in the $L_{\gamma}$--$\Gamma$ plot (shown in the left panel of Figure~\ref{Fig6})
are found. As can be seen, they are more luminous than RGs, a fact 
seen in the comparison of the two sub-classes of AGNs in the Fourth LAT 
AGN Catalog (4LAC-DR3; \citealt{Ajello+22}):
the mean redshift of blazars being $\sim$0.8 while the mean redshift of 
RGs being $\sim$0.05. Also although with large uncertainties on their
\gr\ photon
indices, the three sources appear harder than most of NLSy1 galaxies. 
FSRQs and BL Lacs can be relatively well separated in this plot 
(\citealt{Ghisellini+09}; \citealt{Abdo+09}; \citealt{Ackermann+15}; 
\citealt{Chen18}), as FSRQs are more luminous with
softer emission while BL Lacs are comparably the opposite.
\citet{Chen18} have obtained a criterion line 
($\Gamma = -0.127 \log L_{\gamma} + 8.18$) to separate these two 
sub-types of blazars with a success rate of 88.6\%.
Following this criterion, we may classify 4FGL J0502.6+0036 as a BL Lac, 
4FGL J1708.2+5519 as an FSRQ, and 4FGL J1055.9+6507 as either. 
Additionally, blazars and RGs may also be separated in a
$L_{\gamma}$--$L_{\rm 1.4GHz}$ plot. We obtained the 1.4\,GHz radio 
data from the  NASA/IPAC Extragalactic Database\footnote{\url{https://ned.ipac.caltech.edu/}} (NED), 
and performed a linear fitting between $\log L_{\gamma}$ and 
$\log L_{\rm 1.4GHz}$ for blazars and RGs, respectively. 
For blazars, we obtained a relationship of 
$\log L_{\gamma} = (0.80\pm0.01) \log L_{\rm 1.4GHz} + (11.9\pm0.4)$, 
which is close to the fitting result reported by \citet{Wu+24}. For RGs, 
the relationship we obtained was 
$\log L_{\gamma} = (0.64\pm0.08) \log L_{\rm 1.4GHz} + (17.1\pm3.2)$.
These two relationships are shown in the right panel of Figure~\ref{Fig6}.
As can be seen, the three sources are among blazars and lie close to 
the relationship of blazars. 

Followup optical spectroscopy and X-ray observations of the three sources
are warranted. The first studies can obtain the spectral features that
would further identify them as a BL Lac or FSRQ, or 
an extreme RG or NLSy1 galaxy and possibly determine their redshifts. The latter
will help obtain their X-ray emission properties that help better their
SEDs. In addition, this work exemplarily demonstrates a procedure for 
identification of \fermi-LAT unidentified sources, given the advent of rich
data from large, multi-wavelength surveys. A potential radio counterpart 
will help lower the several-arcminute positional uncertainties 
of \fermi-LAT sources to
a region of a few arcseconds, and multi-wavelength variability analysis
will identify the likely counterpart.

\normalem
\begin{acknowledgements}
This work was based on observations obtained with the Samuel Oschin Telescope
48-inch and the 60-inch Telescope at the Palomar Observatory as part of
the Zwicky Transient Facility project. ZTF is supported by the National
Science Foundation under Grant No. AST-2034437 and a collaboration including
Caltech, IPAC, the Weizmann Institute for Science, the Oskar Klein Center
at Stockholm University, the University of Maryland, Deutsches
Elektronen-Synchrotron and Humboldt University, the TANGO Consortium of
Taiwan, the University of Wisconsin at Milwaukee, Trinity College Dublin,
Lawrence Livermore National Laboratories, and IN2P3, France. Operations are
conducted by COO, IPAC, and UW.
This publication makes use of data products from the Near-Earth Object Wide-field Infrared Survey Explorer (NEOWISE), which is a joint project of the Jet Propulsion Laboratory/California Institute of Technology and the University of Arizona. NEOWISE is funded by the National Aeronautics and Space Administration.
This research has made use of the SIMBAD database, operated at CDS, Strasbourg, France. This research has made use of the NASA/IPAC Extragalactic Database, which is funded by the National Aeronautics and Space Administration and operated by the California Institute of Technology. This work uses data obtained from Inyarrimanha Ilgari Bundara / the Murchison Radio-astronomy Observatory. We acknowledge the Wajarri Yamaji People as the Traditional Owners and native title holders of the Observatory site. CSIRO’s ASKAP radio telescope is part of the Australia Telescope National Facility (https://ror.org/05qajvd42). Operation of ASKAP is funded by the Australian Government with support from the National Collaborative Research Infrastructure Strategy. ASKAP uses the resources of the Pawsey Supercomputing Research Centre. Establishment of ASKAP, Inyarrimanha Ilgari Bundara, the CSIRO Murchison Radio-astronomy Observatory and the Pawsey Supercomputing Research Centre are initiatives of the Australian Government, with support from the Government of Western Australia and the Science and Industry Endowment Fund. This paper includes archived data obtained through the CSIRO ASKAP Science Data Archive, CASDA (http://data.csiro.au).

We thank the referee for helpful suggestions.
This research is supported by the Basic Research Program of Yunnan 
	Province (No. 202201AS070005), the National Natural Science Foundation 
	of China (12273033), and the Original
Innovation Program of the Chinese Academy of Sciences (E085021002).
S.J. acknowledges the support of the science research program for graduate students of Yunnan University (KC-23234629).

\end{acknowledgements}

\bibliographystyle{raa}
\bibliography{bla}

\appendix
\section{Long-term $\gamma$-ray light curves}
\label{sec:lc}

\begin{figure}[htbp]
  \begin{minipage}[t]{0.495\linewidth}
  \centering
   \includegraphics[width=74mm]{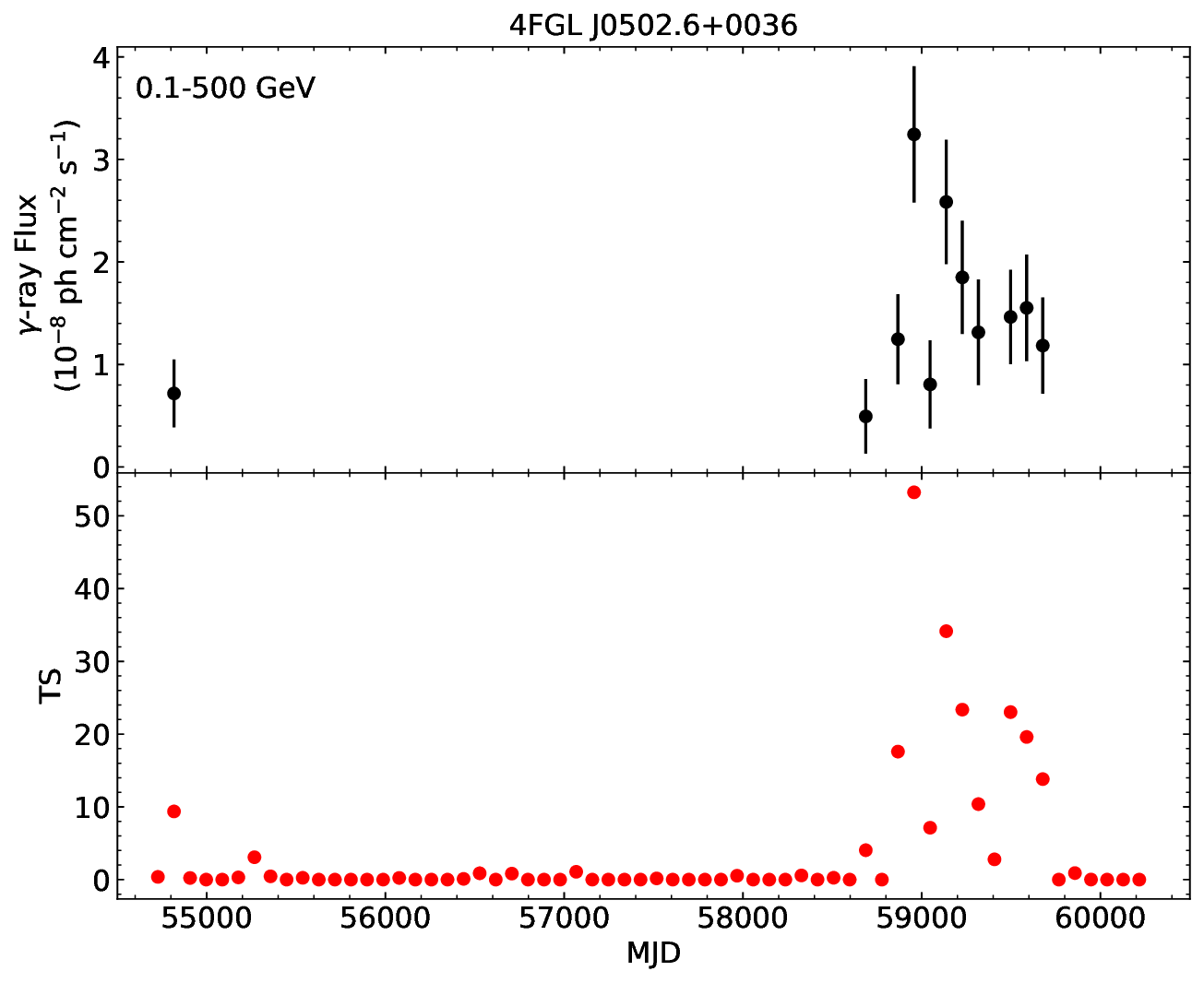}
  \end{minipage}%
  \begin{minipage}[t]{0.495\textwidth}
  \centering
   \includegraphics[width=75mm]{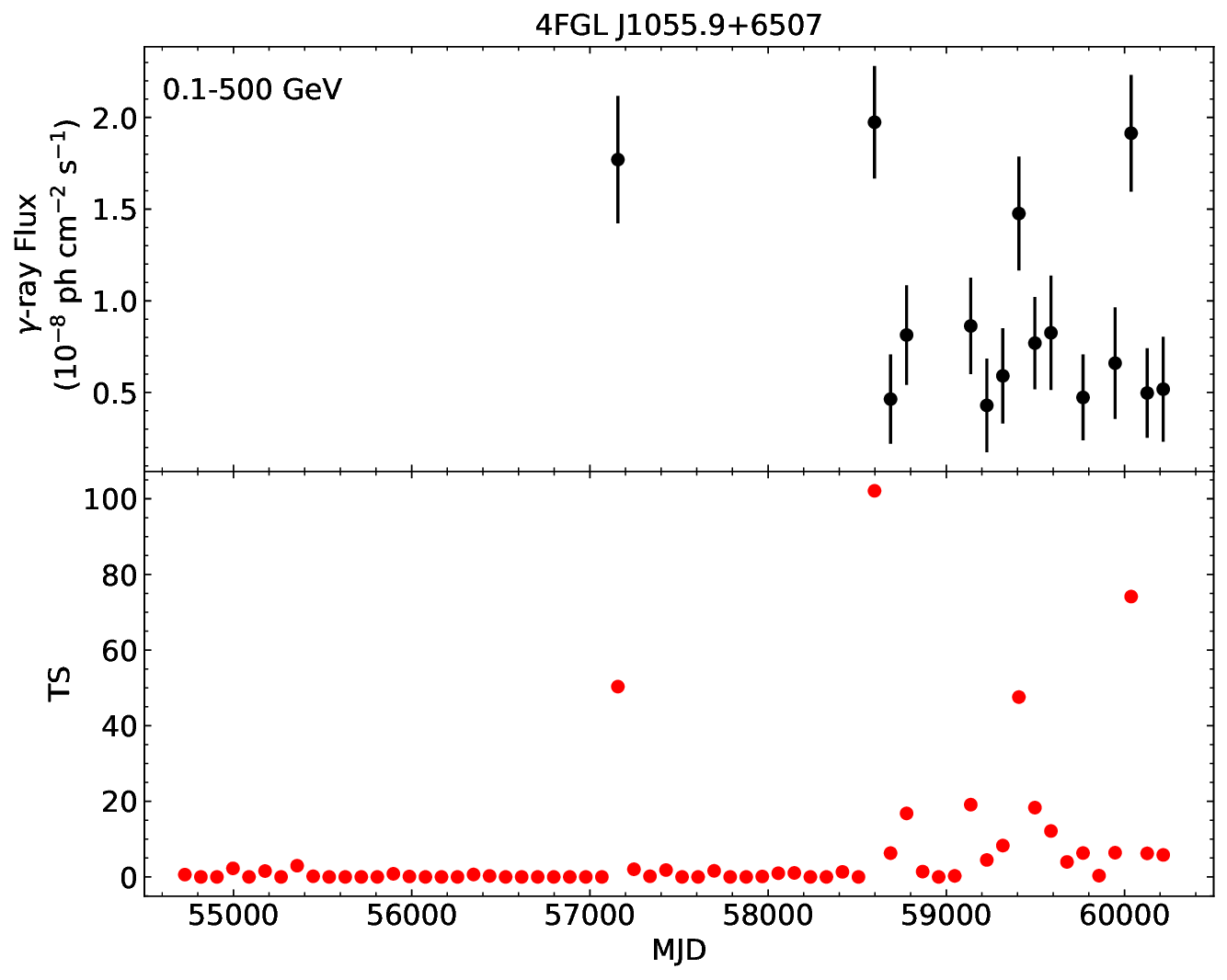}
  \end{minipage}%
  \qquad
   \begin{minipage}[t]{1\textwidth}
  \centering
   \includegraphics[width=75mm]{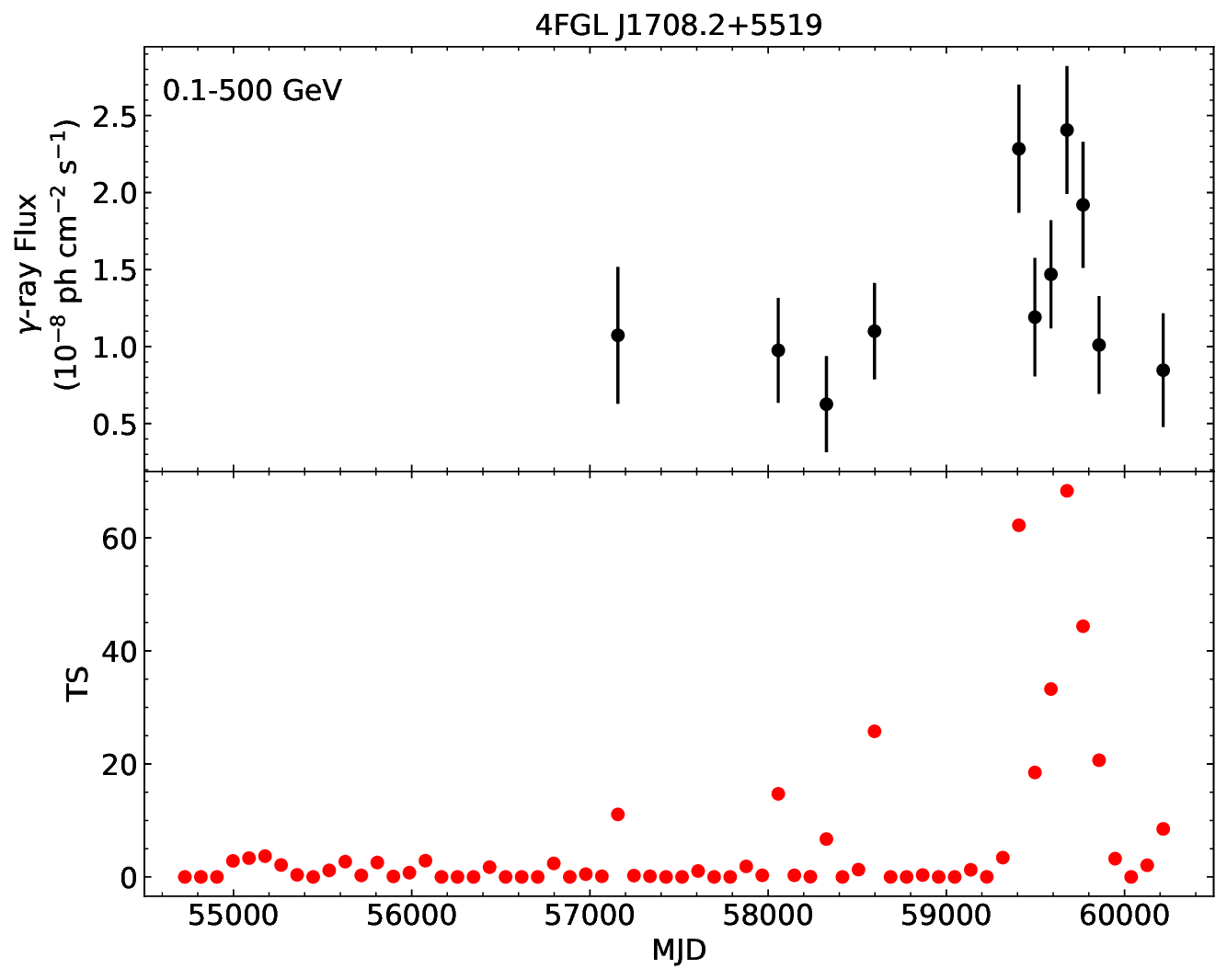}
  \end{minipage}%
	\caption{\label{fig:lc}{90-day binned light curves ({\it top}) and 
	TS curves ({\it bottom}) in 0.1--500 GeV for 4FGL J0502.6+0036, 4FGL J1055.9+6507,
	and 4FGL J1708.2+5519+5519. Fluxes with TS$\geq$4 are kept in the light curves.} }
\end{figure}

\end{document}